# How Soil Organic Matter Composition Controls Hexachlorobenzene-Soil-Interactions: Adsorption Isotherms and Quantum Chemical Modelling

*Ashour A. Ahmed\* [a,c,d], Oliver Kühn [a,d], Peter Leinweber [b,d]*

[a] University of Rostock, Institute of Physics, D-18051 Rostock, Germany

[b] University of Rostock, Soil Science, D-18051 Rostock, Germany

[c] University of Cairo, Faculty of Science, Department of Chemistry, 12613 Giza, Egypt

[d] University of Rostock, Department of Life, Light and Matter, Interdisciplinary Faculty, D-18051 Rostock, Germany

ashour.ahmed@uni-rostock.de

oliver.kuehn@uni-rostock.de

peter.leinweber@uni-rostock.de

**ABSTRACT**

Hazardous persistent organic pollutants (POPs) interact in soil with the soil organic matter (SOM) but this interaction is insufficiently understood at the molecular level. We investigated the adsorption of hexachlorobenzene (HCB) on soil samples with systematically modified SOM. These samples included the original soil, the soil modified by adding a hot water extract (HWE) fraction (soil+3 HWE and soil+6 HWE), and the pyrolyzed soil. The SOM contents increased in the order pyrolyzed soil < original soil < soil+3 HWE < soil+6 HWE. For the latter three samples this order was also valid for the HCB adsorption. The pyrolyzed soil adsorbed more HCB than the other samples at low initial concentrations, but at higher concentrations the HCB adsorption became weaker than in the samples with HWE addition. This adsorption behaviour combined with the differences in the chemical composition between the soil samples suggested that alkylated aromatic, phenol, and lignin monomer compounds contributed most to the HCB adsorption. To obtain a molecular level understanding, a test set has been developed on the basis of elemental analysis which comprises 32 representative soil constituents. The calculated binding free energy for HCB with each representative system shows that HCB binds to SOM stronger than to soil minerals. For SOM, HCB binds to alkylated aromatic, phenols, lignin monomers, and hydrophobic aliphatic compounds stronger than to polar aliphatic compounds confirming the above adsorption isotherms. Moreover, quantitative structure-activity relationship (QSAR) of the binding free energy with independent physical properties of the test set systems for the first time indicated that the polarizability, the partial charge on the carbon atoms, and the molar volume are the most important properties controlling HCB-SOM interactions.

# 1. INTRODUCTION

Persistent organic pollutants (POPs) comprise an environmentally hazardous compound class that resists chemical, biological, and photolytic degradation in the environment (Ritter et al., 2007). They are ubiquitously distributed in the environment having long life times, which can be several days in the atmosphere and years or decades in soil/sediment (Jones and de Voogt, 1999). In aqueous systems and soil, POPs preferentially move into the solid fraction because of their hydrophobicity. In soil, POPs can be taken up by plant roots and/or adsorbed on soil constituents, especially on soil organic matter (SOM). Binding of POPs to SOM is influenced by several factors, including physical and chemical properties of the pollutant, moisture and chemical composition of soil, and the type and strength of the interactions between the pollutant and the reactive soil surfaces (Senesi and Loffredo, 2008). Interactions of hydrophobic pollutants with soil have been studied by adsorption experiments, which usually show an initial rapid and a following slower stage (Chen et al., 2004; Weber et al., 1991). The most common mechanism for the slow stage is diffusion into the SOM (Chiou et al., 1993).

Since adsorption experiments yield only information, which can be correlated statistically to soil properties, computational chemistry is a promising approach to develop an atomistic understanding of the binding of POPs to soil (Gerzabek et al., 2001; Schaumann and Thiele-Bruhn, 2011). For instance, Kubicki, Sparks and coworkers modelled processes at mineral surfaces such as surface complexation with a variety of compounds (Kubicki et al., 2007; Kwon and Kubicki, 2004; Paul et al., 2006; Tribe et al., 2006; Zhu et al., 2009). Other atomistic simulation studies covered, e.g., the binding of polycyclic aromatic hydrocarbons to soot (Kubicki, 2006) and the dynamics of phenol-water (Lock and Skipper, 2007) or salt (Jardat et al., 2009) solutions at clay surfaces. Lischka and coworkers studied the effect of protonation, deprotonation, and dehydroxylation of different reactive sites on a goethite model surface (Aquino et al., 2008) as well as adsorption of polycyclic aromatic hydrocarbons on such a surface (Tunega et al., 2008, 2009). Furthermore, they investigated interactions between a 2,4-dichlorophenoxy acetic acid herbicide and various functional groups (Aquino et al., 2007) and the role of hydrogen bonds in stabilizing poly(acrylic acid) oligomer structures mimicking humic acid (Aquino et al., 2008, 2009). There are different opinions about the principal structural organization of humic substances and SOM, i.e. macromolecular vs. supramolecular structure (Piccolo, 2002; Schaumann, 2006; Sutton and Sposito, 2005). The perhaps most complex polymeric-type, effective atomistic model of SOM has been developed by Schulten and coworkers on the basis of bio- and geochemical, NMR-spectroscopic and mass spectrometric analyses (Schulten, 2002; Schulten and Leinweber, 2000; Schulten and Schnitzer, 1995, 1997). Polymer-like modelling of SOM could be critizised because of the huge number of possibilities for combining all of SOM compounds and functional groups together into a single macromolecule. Therefore, modelling SOM by separate representative systems covering most relevant functional groups as well as analytically determined compound classes is an alternative, which is followed here for the first time using a large test set.

The objective of the present work is to study the interaction of HCB, one of the most dangerous important POPs (Jones and de Voogt, 1999), with SOM experimentally and theoretically. For linking the experimentally observed HCB adsorption by samples with systematically changed SOM composition (Ahmed et al., 2012) a test set of some representative species of the SOM building blocks as well as different functional groups is developed. Then this test set is used to study the interaction of HCB with each representative system computationally by calculating the binding free energy between HCB and each representative system employing density functional theory (DFT). We hypothesize that an improved atomistic understanding of the HCB-SOM-interaction can be derived from correlating the binding free energy with the molecular properties of the representative systems using quantitative structure-activity relationships (QSAR) (Nantasenamat et al., 2010).

## 2. MATERIALS AND METHODS

### 2.1. Soil Samples

The original soil sample was collected from the unfertilized rye plot of the long-term Eternal Rye Cultivation experiment at Halle (Saale), Germany. Controlled experimental modification of SOM in this soil sample was established by changing the polarity character of SOM in two different ways (Ahmed et al., 2012). Firstly, a hot water extract (HWE) fraction, containing mostly polar functional groups such as carbohydrates, N-containing compounds, and peptides, was removed from the original soil sample. Then, this HWE was added into different samples of the same original soil in two different amounts. These amounts were three and six times the HWE content in the original soil sample, and produced the samples soil+3 HWE and soil+6 HWE. This procedure enriched the SOM in mostly polar oxygen-containing functional groups (Ahmed et al., 2012). Secondly, thermal heating by *off-line* pyrolysis at 600 °C of the original soil sample was performed in order to decrease its polar character. These soil samples were characterized using a multi-methodological approach combining elemental analysis, pyrolysis-field ionisation mass spectrometry (Py-FIMS), and C and N $K$-edge X-ray absorption near-edge structure spectroscopy (XANES). These analyses provided evidence for systematically altered SOM compositions (Ahmed et al., 2012) which probably affect the binding of HCB to SOM.

### 2.2. Adsorption of Hexachlorobenzene

For the adsorption experiments, a stock solution of 100 µg/ml HCB (CAS number 118-74-1, Sigma-Aldrich) was prepared in *n*-hexane. Different initial HCB concentrations were prepared by dilution of this HCB stock in $CaCl_2$ solution.

One gram of each air-dry soil sample (original soil, soil+3 HWE, soil+6 HWE, and pyrolyzed soil) was mixed with 30 ml of different initial concentrations of HCB in Teflon tubes. These HCB concentrations are 0.25, 0.50, 0.75, 1.00, 2.00, 3.00, 4.00, and 5.00 µg/ml. They were prepared in 0.01 M CaCl$_2$ solution using the HCB stock solution in *n*-hexane. To suppress microbial activity, 100 µl of 100 mg/l NaN$_3$ was added to each suspension. In addition, two blank measurements were also processed at the same time. One of them with 5.00 µg/ml HCB without soil sample and the other with soil sample at 0.00 µg/ml HCB. All the adsorption and blank measurements were done in duplicates. The suspensions were shaken at 22 revolutions per minute using a special overhead rotator (GFL overhead rotator 3040) for 24 hours. Then they were centrifuged for 20 min at 3500 ×g. For HCB analysis, 50 µl of n-hexane layer, containing HCB, was sampled from each tube. HCB concentration was determined by using a G1530A (Agilent Technologies, Santa Clara, USA) gas chromatograph with two parallel capillary columns with different polarities, each equipped with an electron capture detector (ECD). The separation was perform by a 60 m Varian FactorFour capillary column VF-5ms (5% phenylmethyl- and 95% dimethylpolysiloxane) with an inner diameter of 0.25 mm and a film thickness of 0.25 µm and by a J & W capillary column DB-1701P (14% cyanopropylphenyl- and 86% dimethylpolysiloxane) with 60 m length, 0.25 µm film thickness and an inner diameter of 0.25 mm. In a split less mode, 1 µl of HCB sample was injected. HCB concentrations were determined by comparison of the peak height of the analyte with that of HCB standards.

The adsorption isotherms were fitted by using the Freundlich equation,

$$X = K_F \, C_{eq}^n \tag{2.1}$$

where $X$ is amount of adsorbed HCB on the soil sample (given either in µg/g soil or in µg/g total carbon content, $C_{tot}$), $K_F$ is Freundlich unit capacity factor, $n$ is Freundlich exponent, and $C_{eq}$ is the HCB equilibrium concentration in µg/ml. Various studies reported the strong relationship between the total organic carbon in the soil and the mobility of the pesticides (Jodeh et al., 2009). For this reason, the mount of the adsorbed HCB on the different soil samples was normalized to the total carbon content in the soil sample. Notice that Freundlich model assumes that the adsorption enthalpy depends on the amount of adsorbed HCB. In the limit of small $X$ where the adsorption enthalpy should not depend on $X$ one could describe the isotherm by a Langmuir equation as well, i.e.

$$X = X_{max} \frac{K_L \, C_{eq}}{1 + K_L \, C_{eq}} \tag{2.2}$$

where $X_{max}$ is the maximum amount of adsorbed HCB on the soil samples, which is required to have a complete saturation of all binding sites and $K_L$ is the equilibrium Langmuir constant.

## 2.3. Soil Organic Matter Modelling

To reduce the problems arising from polymer-like modelling of SOM or modelling of SOM by few numbers of functional groups, we have developed a new approach. Specifically, we have modelled the SOM by separate representative systems covering almost all functional groups as well as analytically determined compound classes of SOM. Development of a SOM model for studying HCB-SOM-interaction has been based on detailed molecular analyses by Py-FIMS and XANES at the C and N *K*-edges (Ahmed et al., 2012). The SOM building blocks identified by Py-FIMS include phenols + lignin monomers (PHLM), alkyl aromatics (ALKY), carbohydrates (CHYDR), heterocyclic nitrogen containing compounds (NCOMP), peptides (PEPTI), lipids, alkanes, alkenes, bound fatty acids, and alkyl monoesters (LIPID), and lignin dimers (LDIM). Hence, in the test set of representative SOM compound classes and functional groups (Figure 1) PHLM are modeled by phenol (**22**), catechol (**26**), and 3,4,5-trimethoxy cinnamic acid (**30**) (lignin monomer). ALKY are modelled by benzene (**17**), methylbenzene (**18**), and ethylbenzene (**24**). Moreover bicyclic aromatic compounds, like naphthalene (**28**) and ethylnaphthalene (**29**), are added to study effect of increasing number of aromatic rings. CHYDR are represented by the most abundant monomer glucose in the open (**2**) and cyclic (**15**) forms. PEPTI are modelled by the main abundant monomer glycine (the HCB-glycine complex has two energetically equivalent configurations (**9**) and (**10**)) and hexa-glycine (**27**). NCOMP are represented by ethylnitrile (**3**), five- and six-membered heterocyclic compounds pyrrole (**16**) and pyridine (**13**), respectively. Acetic acid (**6**) is modelled as carboxylic acid in the free fatty acids. LIPID, alkane, alkene compounds are represented by short chain alkane (**12**) and conjugated alkene (**14**), and long chain alkane (**23**) and conjugated alkene (**20**). Effect of sterols can be understood from including of the hydroxyl group in methanol (**4**), alkanes (**12**,**23**), and alkenes (**14**,**20**). Moreover, based on recent functional groups analysis (Ahmed et al., 2012), we have also added to our model set carbonyl such as acetamide (**1**), acetaldehyde (**5**), dimethylketone (**7**), and methylacetate (**11**); amine like methylamine (**8**), protonated methylamine (**21**), and aniline (**25**); and quinone (**19**). In addition, coronene (**31**) and silicon hydroxide trimer (**32**) are added to study the effect of pyrolysis products on binding of HCB to the soil.

**Figure 1**

## 2.5. Quantum Chemical Calculations

Initial geometries of complexes of these test set compounds with HCB were constructed by selecting the expected preferential binding situations for each complex. Full geometry optimization was performed for all individual species (HCB and each test set system) and all complexes in the gas phase. In case this resulted in more than one configuration the most stable one has been selected. An exception is the HCB-glycine complex, which has two equivalent configurations (**9**, **10**). The binding energies of HCB to the test set compounds in these complexes were calculated as the difference between the total energies of the complex and the individual molecules.

$$E_{B_i} = E_{(HCB-i)complex} - (E_{HCB} + E_i) \qquad (2.3)$$

where, $E_{B_i}$ is the binding energy of HCB to the compound $i$, $E_{(HCB-i)complex}$ is the energy of the complex of HCB with the compound $i$, $E_{HCB}$ is energy of HCB, and $E_i$ is energy of the compound $i$.

The interaction of HCB with the test set has been studied by quantum chemical DFT calculations. Here, the Becke, three-parameter, Lee-Yang-Parr hybrid functional (B3LYP) (Becke, 1988; Lee et al., 1988) has been used together with a 6-311++G(d,p) basis set. Dispersion energies are accounted for by employing the empirical D3 correction due to Grimme and coworkers (Grimme et al., 2011). Ability of this quantum mechanical level of theory combined with the dispersion correction D3 to describe this interaction type was checked versus standard methods such as MP2 and CCSD (Ahmed et al., submitted). In addition, effect of the basis set superposition error (BSSE) (Jansen and Ros, 1969) has been neglected in case of DFT-D3 due to the binding energies in case of the uncorrected DFT-D3 from BSSE are closer to those obtained by corrected MP2 than those in case of the corrected DFT-D3 from BSSE (Ahmed et al., submitted). Since the soil solution, which is mainly composed of water, is an important factor controlling this interaction, it was simulated by a continuum solvation approach. Solvation by water has been included within the conductor-like screening model (COSMO) (Schäfer et al., 2000). Using COSMO, full geometry optimization was performed for all individual species and their complexes. The binding free energy of HCB to each test set compound in these complexes was calculated. All calculations have been performed using the TURBOMOLE program package (Ahlrichs et al., 1989).

Moreover, QSAR analysis has been done to correlate the calculated binding free energy ($E_B$) of HCB to SOM representative systems with the appropriate physical parameters governing this interaction. The isotropic polarizability ($P_1$), quadrupole moment ($P_2$), sum of the partial charges on the carbon atoms ($P_3$), sum of the partial charges on the nitrogen atoms ($P_4$), molecular-mass ($P_5$), and molar volume ($P_6$) of the test set systems are used as descriptors. These physical properties are correlated to the binding free energies via the following equation.

$$E_B = C_0 + C_1 P_1 + C_2 P_2 + C_3 P_3 + C_4 P_4 + C_5 P_5 + C_6 P_6 \qquad (2.4)$$

The coefficients $C_0$, $C_1$, $C_2$, $C_3$, $C_4$, $C_5$, and $C_6$ were determined using multiple-linear regression. In addition, some statistical parameters were calculated such as sum of squares due to the error (SSE), sum of squares due to the regression (SSR), sum of total squares (SST), mean of squares due to the error (MSE), mean of squares due to the regression (MSR), and mean of total squares (MST). Moreover, $R^2$ (which is equal to SSR/SST) and adjusted $R^2$ (which is equal to 1-MSE/MST) which are proportional to the total variation, and $F_{statistic}$ (which is equal to MSR/MSE) which measures significance of the model describing the data were calculated.

## 3. RESULTS AND DISCUSSION

### 3.1. HCB Adsorption

The adsorption of HCB on the soil samples increased upon addition of HWE resulting in the order original soil < soil+3 HWE < soil+6 HWE (Figure 2). This is valid for adsorbed HCB concentrations normalized to the total soil mass (Figure 2 A) and the total carbon content in the soil (Figure 2 B). Adsorption of HCB on the pyrolyzed soil sample exceeded that of soil+6 HWE at low initial concentrations when normalized to the total soil mass. At increasing HCB concentrations, the adsorption of HCB on the pyrolyzed soil sample becomes smaller (Figure 2 A). The same is valid over a larger range of HCB concentrations when normalized to the total carbon content in the soil (Figure 2 B).

**Figure 2**

The fitted isotherms to Freundlich and Langmuir equations are shown in Figure 3 and Figure 4, respectively. The fitted parameters of adsorption isotherms yielded comparable squared correlation coefficients ($r^2$) close to one for both equations (Table 1). Considering the Freundlich model, the form of the obtained isotherms supports the prevailing sorption mechanism of a given substance in the system. Nonlinearity of Freundlich exponents ($n$: 0.56 to 0.80) in all soil samples indicates that the sorption mechanism is dominated by adsorption and not absorption (Toul et al., 2003). Furthermore the exponent $n$ indicates the diversity of the free energies associated with adsorption of HCB on a heterogeneous surface. The $n < 1$ for all soil samples indicates that upon increasing the HCB concentration the binding is reduced, i.e. the binding free energies decrease. The only significant difference is observed for the pyrolyzed sample ($n = 0.56$). For this reason, HCB adsorption on the pyrolyzed sample is described slightly better by the Langmuir than by the Freundlich equation. The order of adsorption isotherms, original soil < soil+3 HWE < soil+6 HWE (Figure 2) is reflected by the increase of $K_F$ and $X_{max}$ in Table 1. The Freundlich isotherm parameters for the soil+3 HWE and soil+6 HWE are similar to those obtained for samples from a red and a paddy soil in China, respectively (Gao and Jiang, 2010). Although these authors did not report the carbon content in the soil we could explain this by similarity in organic matter contents because the mineralogy must be completely different from the present soil.

**Figure 3 & Figure 4**

In the Py-FIMS characterization of SOM quality the absolute ion intensity (*AII*) of each compound class increased in the order original soil (*AII*1) < soil+3 HWE (*AII*2) < soil+6 HWE (*AII*3), indicating a contribution of all SOM constituents to HCB adsorption (Table 2). In order to find out how SOM constituents differ in their contribution to HCB adsorption, we introduce the amount of adsorbed HCB on original soil, $X_1$, on soil+3 HWE, $X_2$, and on soil+6 HWE, $X_3$. For a given equilibrium concentration the difference in the adsorbed concentrations between the original soil and soil+3 HWE ($X_2$-$X_1$) is greater than that between the soil+3 HWE and soil+6 HWE ($X_3$-$X_2$) (Figure 2). For comparison of ($X_2$-$X_1$) and ($X_3$-$X_2$) with the molecular SOM composition we denote the differences in *AII* for each compound class *ΔAII*$_1$=*AII*2-*AII*1, and *ΔAII*$_2$=*AII*3-*AII*2. For LIPID and LDIM *ΔAII*$_1$ < *ΔAII*$_2$ but for PHLM, ALKY, CHYDR, NCOMP, and PEPTI *ΔAII*$_1$ > *ΔAII*$_2$. This suggests that the later compound classes are more likely to explain the above differences in HCB adsorption among samples enriched in hot water extracted organic matter. Especially PHLM, and ALKY, having the largest *ΔAII*$_1$:*ΔAII*$_2$ values, might contribute to the binding of HCB to SOM more significantly than the other compound classes. This will be tested by the binding energy calculations in section 3.2.

The low total ion intensity in Py-FIMS of the pyrolyzed soil did not allow determining the *AII* of each compound class (Ahmed et al., 2012). However, it is well known that pyrolysis in general decreases the amount of SOM and increases the proportion of unsaturated, substituted aromatic, heterocyclic, and aliphatic nitrile compounds, besides producing charcoal (Ahmed et al., 2012; Kiersch et al., 2012a, 2012b). Hence it is reasonable to assume that such changes upon pyrolysis process are responsible for the behaviour shown in Figure 2.

**3.2. Quantum Chemical Modelling**

The gas phase equilibrium geometries of the test set complexes are collected in Figure 5. The numbering of these complexes of the representative systems is according to increasing binding energies of HCB to the representative systems (except the inorganic silicon hydroxide) in case of gas phase. Figure 5 shows that HCB interacts through its positively charged hydrophobic ring center with the negatively charged center of most modelled systems. There are three exceptions (acetamide (**1**), glucose in the open form (**2**), and charged methylamine (**8**)) that bind HCB in a different way. In complexes of these systems, two chlorine atoms in HCB interact with one or two H atom(s). Furthermore, Figure 5 clearly shows that there are no covalent bonds formed between HCB and the SOM model set, i.e. binding is due to dispersion interaction except for the charged system (**21**).

**Figure 5**

The binding energies in the gas phase for the test set complexes (Figure 6) indicate that aromatic compounds (**13**, **16-19**, **22**, **24-26**, and **28-30**) bind HCB stronger than aliphatic compounds (**1-12**, **14**).

This can be explained by the type and strength of the interaction center. For polar-aliphatic compounds, the centers of interaction are the partially negatively charged atoms, while for non-polar aliphatic compounds (like alkanes and alkenes) most of the atoms contribute to the interaction. For aromatic compounds, the centers of interaction are the partially negatively charged aromatic rings. Thus, it can be concluded that the binding energy increases with the subjected surface area for the interaction. This also implies that binding energies for HCB with long chain alkanes and alkenes are comparable to that of aromatic compounds. A more detailed inspection of Figure 6 reveals that for aliphatic compounds HCB binds in the order saturated long chain hydrocarbon (**23**) ~ unsaturated long chain hydrocarbon (**20**) > unsaturated short chain hydrocarbon (**14**) ~ saturated short chain hydrocarbon (**12**) > amine functional group (**8**) > carbonyl functional group (**5-7**) > alcohol functional group (**4**) > nitrile functional group (**3**). In case of aromatics we find that HCB binds in the order aniline (**25**) > ethylbenzene (**24**) > phenol (**22**) > methylbenzene (**18**) > benzene (**17**). Further, HCB binds to carbohydrates (**15**) (modelled by glucose) and peptides (**27**) (modelled by hexaglycine) within the aromatic's binding range. Due to the above mentioned functional group effect, binding is stronger to peptides than to carbohydrates. For the same reason HCB binds alkylated aromatic compounds (**17**, **18**, **24**) stronger than heterocyclic ones (**13**, **16**). Within the aromatic compounds, HCB binds the polycyclic aromatic rings (like the substituted (**29**) and non-substituted (**28**) naphthalene) stronger than monocyclic aromatic rings (like the substituted (**18**, **24**) and non-substituted (**17**) benzene). Note that despite HCB binds to naphthalenes stronger than to benzenes, the interaction with HCB exceeds that of naphthalenes if benzene is substituted by a strong electron donating functional group (like the lignin monomer (**30**)).

**Figure 6**

The COSMO calculations show that solvation does not affect significantly the spatial configuration of HCB-test set complexes. The root-mean square deviation between the conformers in the gas phase and the corresponding ones in solution is less than 0.05 Å for most of these complexes. Only in the case of HCB-acetic acid and HCB-methylacetate complexes, the systems were rotated by 90° to make the planes containing the system and HCB parallel to each other. The calculated binding free energies show that solvation decreases the binding energy for all HCB-test set complexes from gas phase to solution as shown in Figure 6. This is due to stabilization of the inorganic species as well as the SOM components by water. Nevertheless, the overall picture remains almost unchanged, i.e. HCB binds to both aromatic and non-polar aliphatic compounds more than to polar aliphatic compounds. Specifically, HCB binds in the order: substituted polycyclic aromatic compounds like naphthalenes (binding free energy: -14.2 kcal/mol) > lignin monomers (-13.7 kcal/mol) > long chain alkanes (-11.4 kcal/mol) > substituted benzenes with alkyl and amino groups ~ long chain alkenes (-9.6 kcal/mol) > phenols (-9.3 kcal/mol) > short chain alkanes (-7.4 kcal/mol) > five membered heterocyclic ring compounds (-7.1 kcal/mol) >

short chain alkenes ~ esters (-6.6 kcal/mol) > carbohydrates (-6.2 kcal/mol) > peptides (-5.2 kcal/mol) > six membered heterocyclic ring compounds (-4.5 kcal/mol) > polar aliphatic compounds (-2.9 kcal/mol). As a general trend it is observed that the binding free energy decreases with increasing polarity of SOM components. However, the charged amine (**21**), hexaglycine (**27**), and silicon hydroxide trimer (**32**) behave exceptional as they show a strong decrease in binding free energy (marked (I), (II), and (III) respectively in Figure 3). Here the solvation of the positive charge and the highly polar functional groups compensate the other types of the interaction (electrostatic and dispersion).

Figure 6 can be summarized into four points:

1. HCB binds to aromatic and non-polar aliphatic compounds stronger than to polar aliphatic and inorganic compounds.

2. HCB binds polycyclic aromatic compounds stronger than monocyclic aromatic compounds.

3. As the subjected surface area for the interaction increases, binding of HCB increases as well.

4. Solvation reduces the binding energies in all cases especially for polar aliphatic compounds, peptides, and carbohydrates.

In order to establish a correlation between computational and experimental results, the binding free energy values of the different test set complexes were grouped according to their compounds classes (see above). The averaged binding free energies for these compound classes are given in Table 2. Let us denote the total binding free energy ($E_B$) of HCB to original soil, soil+3 HWE, and soil+6 HWE with $E_{B,org}$, $E_{B,3HWE}$, and $E_{B,6HWE}$, respectively. Next we assume that the $E_B$ for HCB to any soil sample is directly proportional to the *AII* and the average binding energy for each compounds class in the soil $\langle E_B \rangle_i$, i.e.

$$E_B \propto \langle E_B \rangle_i * AII_i \tag{3.1}$$

where, the sum runs over all compounds classes.

Since Figures 2, 3, and 4 and Table 2 suggested that important differences might be seen in the relative changes, using (2.1) we consider

$$E_{B,3HWE} - E_{B,org} \propto \sum_i \langle E_B \rangle_i * (AII_{i,3HWE} - AII_{i,org}) = \sum_i \langle E_B \rangle_i * \Delta AII_{1,i} \tag{3.2}$$

$$E_{B,6HWE} - E_{B,3HWE} \propto \sum_i \langle E_B \rangle_i * (AII_{i,6HWE} - AII_{i,3HWE}) = \sum_i \langle E_B \rangle_i * \Delta AII_{2,i} \tag{3.3}$$

Let us further assume that the ratio of change of the amount of adsorbed HCB is proportional to the change in binding energy, i.e.

$$\frac{E_{B,3HWE} - E_{B,org}}{E_{B,6HWE} - E_{B,3HWE}} \propto \frac{X_2 - X_1}{X_3 - X_2} \tag{3.4}$$

Then the ratio of binding energy changes is about 2.8 (Table 2). This in fact is in accord with the observation that $X_2$-$X_1$/$X_3$-$X_2$>1. The largest contribution to $E_{B,3HWE}$-$E_{B,org}$ is due to PHLM and ALKY. This is not outweighted by any of the contribution to $E_{B,6HWE}$-$E_{B,3HWE}$. Thus, we find clear indications that PHLM and ALKY compounds classes are dominating the adsorption behaviour of HCB on the soil samples under study.

To explain the HCB adsorption on the pyrolyzed soil, additionally, coronene (**31**) and a silicate (**32**) segment (silicon hydroxide trimer) were added to the model set to mimic the highly aromatic character of pyrolyzed SOM and soil mineral surfaces, respectively. Taking into account the solvent effect, the binding free energies of HCB to coronene are -17.67 kcal/mol and to the silicate segments -4.49 kcal/mol. First, these values are in accord with the widely accepted view that SOM has a higher impact on adsorption of hydrophobic organic compounds on soil than soil minerals (Holmén and Gschwend, 1996; Karickhoff et al., 1979; Schwarzenbach and Westall, 1981). Second, these data in combination with the fact that the SOM content in the pyrolyzed soil sample is around 75% of that in the original soil sample (Ahmed et al., 2012), explain the behaviour of the adsorption isotherms in Figure 3. The pyrolyzed soil sample is of highly aromatic character and contains unsaturated organic compounds. These compounds bind HCB strongly so that for small HCB concentrations, the pyrolyzed sample will adsorb stronger than the other samples. By increasing HCB concentration, adsorption of HCB on the pyrolyzed soil sample starts to saturate due to the low SOM content. Therefore, with increasing HCB concentration adsorption on the pyrolyzed soil gradually drops below that of the other samples. This argument is in good agreement with the Freundlich exponent *n* for adsorption of HCB on the pyrolyzed soil sample which has the lowest value (0.56) compared to the other soil samples. This indicates that the binding free energy decreases stronger for the pyrolyzed than for the other soil samples. An intuitive picture, which is in accord with the polymer-like SOM models (Schulten and Schnitzer, 1995, 1997), would be that pathways for diffusion through the SOM complex to potential binding sites are blocked with increasing HCB concentration.

### 3.3. Quantitative Activity-Structure Relationship

The binding free energies of HCB-SOM complexes were correlated to different physical properties of the test set systems. This correlation was done based on the quantitative activity-structure relationship (QSAR) as described in Section 2.5. The coefficients of equation (2.4), $C_0$, $C_1$, $C_2$, $C_3$, $C_4$, $C_5$, and $C_6$, were determined using the multiple-linear regression and given in the following equation.

$$E_B = -2.072 - 0.568\, P_1 + 0.180\, P_2 + 1.348\, P_3 - 0.530\, P_4 + 0.014\, P_5 + 0.061\, P_6 \qquad (3.5)$$

The fitted binding free energy in the above equation will be called estimated binding free energy. These estimated binding free energies versus the actual ones are plotted in Figure 7. The fitted parameters of this equation are collected in Table 3. The large value of $R^2$, the small difference between $R^2$ and the adjusted $R^2$, and the larger value of $F_{statistic}$ than critical F indicates to a good and an efficient QSAR equation.

**Figure 7**

Since validity of equation (3.5) has been established, correlations as well as contributions of the different descriptors in absence of the other descriptors can be introduced. QSAR analysis indicates that the most correlated descriptors contributing to the binding free energy are the polarizability (79.7%), sum of the partial charges on carbon atoms (29.9%), and the molar volume of the representative SOM systems (26.4%). Both polarizability and molar volume of the representative SOM systems are negatively correlated to the binding free energy. On the other hand, sum of the partial charges on carbon atoms is positively correlated to the binding free energy. This means that increasing the polarizability and/or the molar volume and/or decreasing sum of the partial charges on carbon atoms of SOM will increase binding of HCB to SOM systems. Moreover, the polarizability is highly positively correlated to the molar volume of SOM systems. The sum of the partial charges on carbon atoms is negatively correlated to that on nitrogen atoms of SOM. The molar volume of SOM has high positive correlation with the molecular mass and the quadrupole SOM systems.

Understanding of nature of HCB-SOM interaction can be highly supported by QSAR. Firstly, since the polarizability of SOM systems is the most predominant property affecting this interaction, this means that the dispersion interaction is the predominant type of interaction. Secondly, importance of the partial charges on carbon atoms in QSAR indicates the important role of type as well as number of carbon atoms on binding of HCB to SOM systems. This confirms our experimental results that the concentration of carbon (Fig. 3) as well as the type of carbon atoms (Tab. 2) determine the adsorption behavior. This becomes obvious in case of the pyrolyzed soil sample compared with the other soil sample. Thirdly, the dependence of the binding free energy on molar volume of SOM agrees with our theoretical suggestion that in case the subjected surface area of SOM increases, the binding energy increases. Moreover, since the partial charges on carbon atoms are negatively correlated with the partial charges on nitrogen atoms, this means that the binding free energy increases as the partial charges on nitrogen atoms decrease. This can be translated into the statement that in case the polarity of SOM system decreases (hydrophobicity increases), its binding to HCB increases.

## 4. Summarizing Discussion

Adsorption of HCB on the pyrolyzed soil, original soil, soil+3 HWE, and soil+6 HWE samples was studied by batch-experiments. The adsorption of HCB increased upon addition of HWE in the order: original soil sample < soil+3 HWE sample < soil+6 HWE sample. At low initial concentrations, adsorption of HCB on the pyrolyzed soil sample exceeded that of soil+6 HWE. The adsorbed concentrations of HCB on the different soil samples were correlated with the absolute ion intensities of the each SOM compound class. This indicated that both PHLM and ALKY are the most important compound classes for adsorption of HCB on soil samples modified by addition of HWE fractions. The unsaturated, substituted aromatic, and heterocyclic compounds besides charcoal are the most effective compounds type controlling adsorption of HCB on the pyrolyzed soil sample.

For a molecular-level understanding of the HCB-SOM interaction, a new SOM model has been developed. The advantage of this model is that we neither make an assumption on an arrangement of functional groups in a supermolecule or polymer (Schulten, 2002; Schulten and Leinweber, 2000; Schulten and Schnitzer, 1995, 1997) nor restrict ourselves to a few functional groups only (Aquino et al., 2007). Moreover studying the interaction of pollutant molecules with a rather large set of functional groups allows to draw conclusions based on statistics, which is closer to the nature of available experimental data. For the particular case of HCB it was found that it interacts through its positively charged hydrophobic ring with the negatively charged centers of the test set molecules. Within a continuum solvation model, HCB binds in the order: substituted polycyclic aromatic > lignin monomers > long chain alkane > substituted benzene ~ long chain alkene > short chain alkane > five membered heterocyclic > short chain alkene ~ ester > carbohydrate > peptide > six membered heterocyclic > polar aliphatic compounds. The estimated binding free energies at the COSMO-B3LYP/D3/6-311++G(d,p) level of the different soil samples were correlated with the corresponding adsorbed HCB concentrations. This showed that PHLM and ALKY compound classes are dominating the adsorption of HCB on the different soil samples. Due to the low SOM content but the high organic character of the pyrolyzed soil sample, it binds HCB stronger than the other soil samples.

Having at hand molecular level information on the different complexes one can identify key parameters that are responsible for the binding interaction. Large binding free energies have been found for HCB and compounds having highly aromatic character and/or unsaturated centers while a small binding free energy between HCB and the modelled inorganic system was observed. In addition, a QSAR analysis showed that polarizability, molecular volume and mass, and charges and percentage of the carbon atoms of SOM systems are the most vital parameters controlling this interaction.

In summary, we have demonstrated a novel approach to the study of pollutant-SOM interaction which combines a host of analytical methods applied to controlled modifications of soil samples (Ahmed et al, 2012) with the development of a large molecular test set and the computational study of its interaction with the pollutant. Key parameters of this interaction can be derived and by comparison with adsorption studies be correlated to the binding of the pollutant. Future applications to other pollutants will serve to validate and further improve the computational model.

**ACKNOWLEDGMENT**

This work was financially supported by the Interdisciplinary Faculty (INF), Department of Life, Light, and Matter, University of Rostock, Germany. The authors would like to thank Dr. Sergei D. Ivanov (Institute of Physics, Rostock University) for his helpful discussion on the quantum chemical calculations.

**Graphical Abstract**

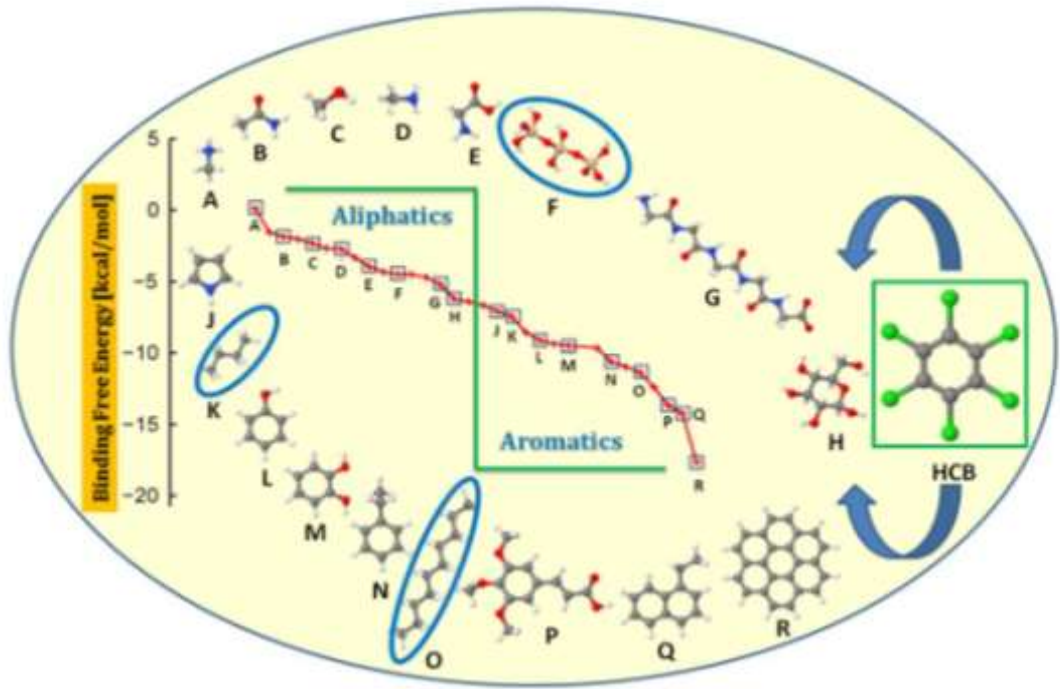

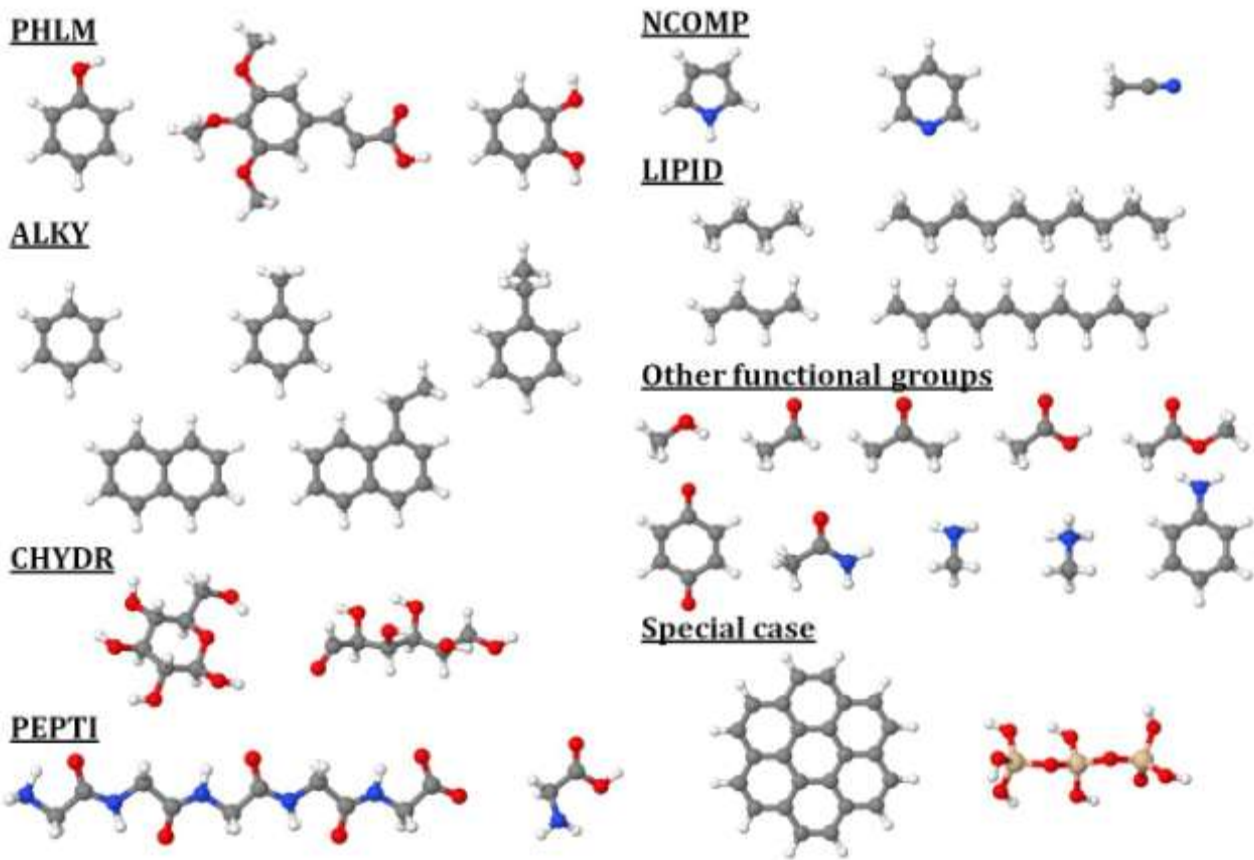

Figure 1. The developed test set for studying the interaction of HCB with different functional groups present in SOM.

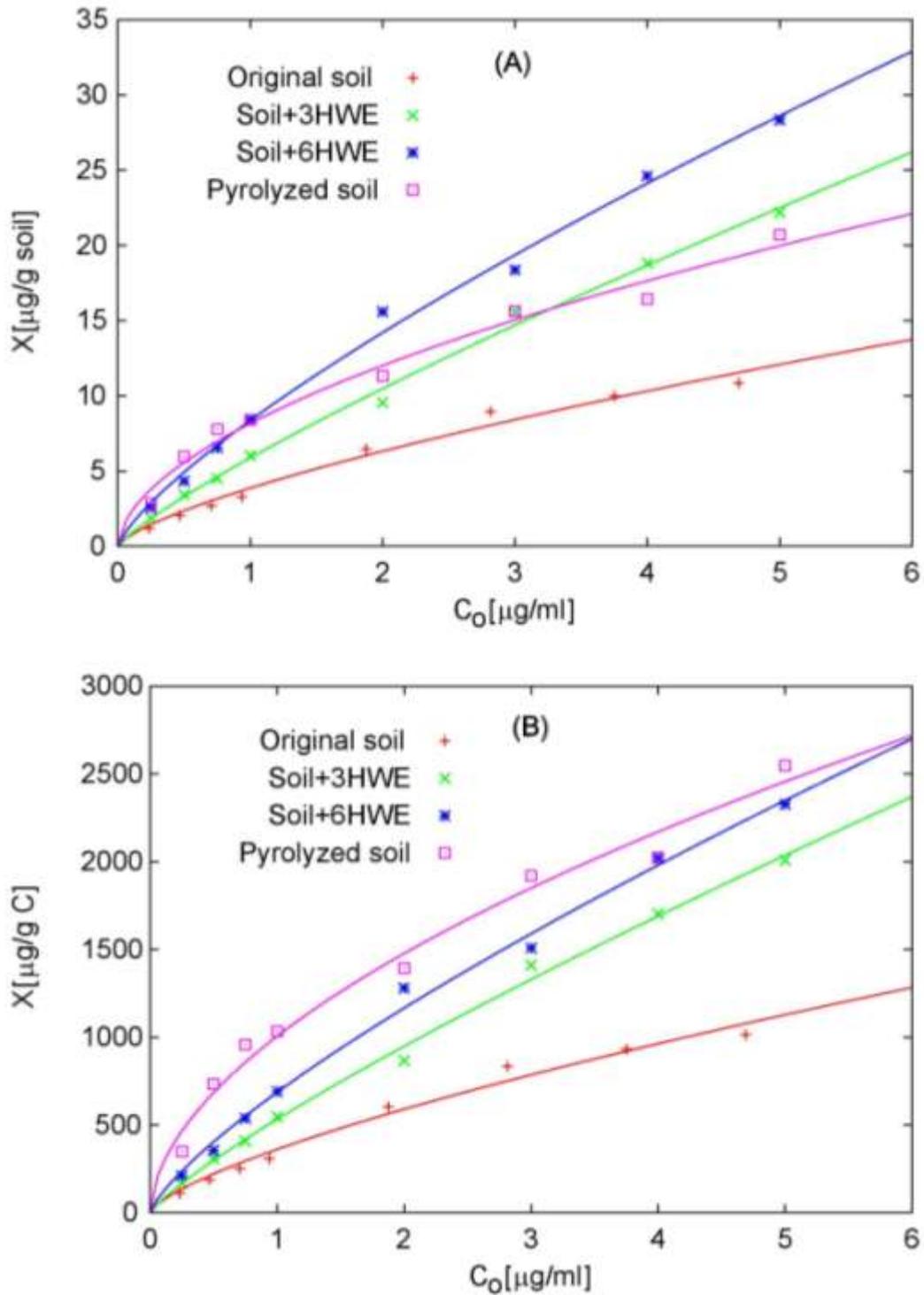

Figure 2. The adsorption isotherms of HCB on the soil samples in which amount of adsorbed HCB normalized to the total soil mass, in μg HCB/g soil, (A) and the total carbon content, in μg HCB/g $C_{tot}$, (B) were plotted versus the corresponding initial HCB concentrations ($C_0$). The lines were, obtained from an exponential correlation, plotted as guide for eye

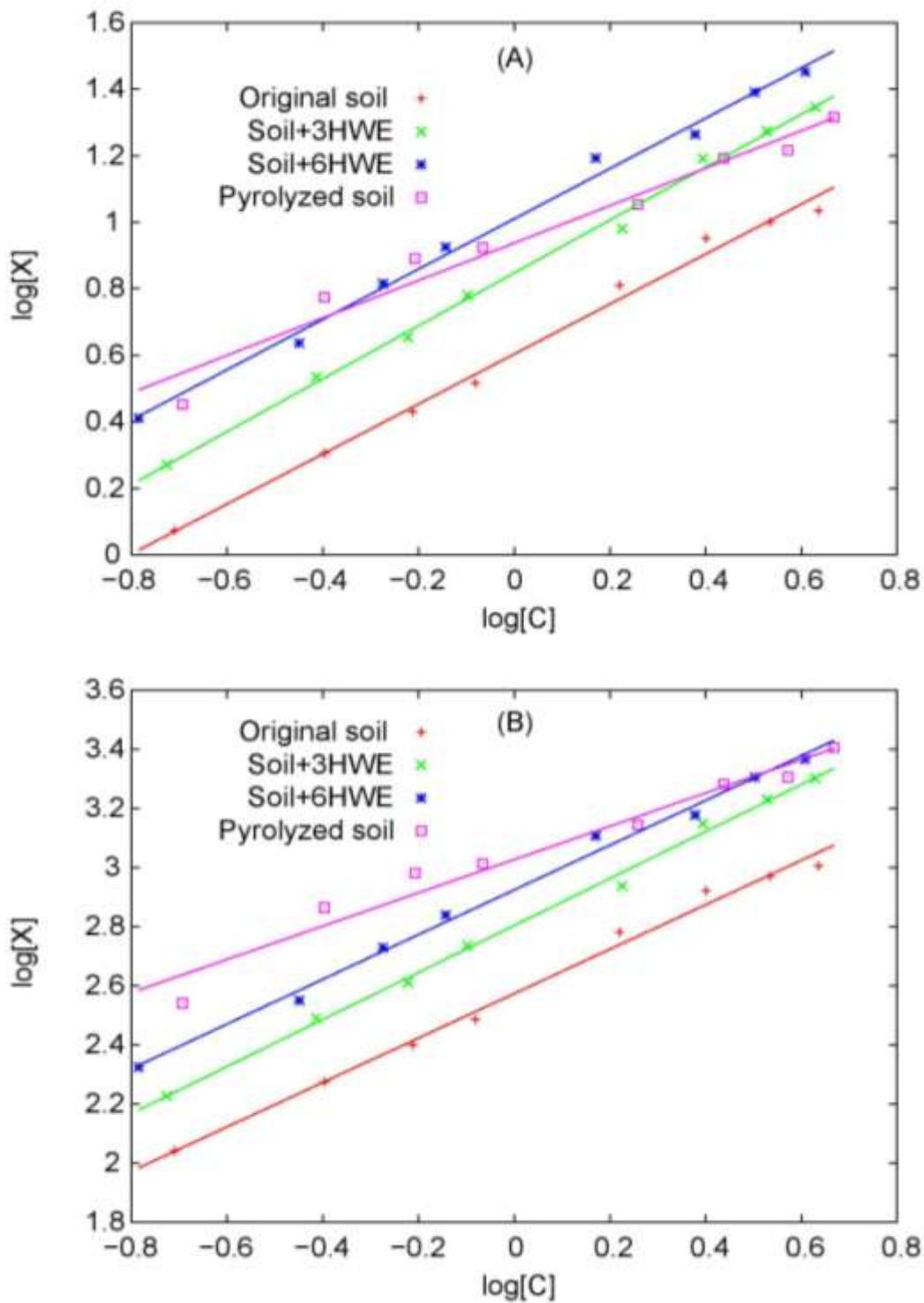

Figure 3. The Freundlich fitted adsorption isotherms of HCB on the soil samples normalized to the total soil mass, in μg HCB/g soil, (A) and the total carbon content, in μg HCB/g $C_{tot}$, (B).

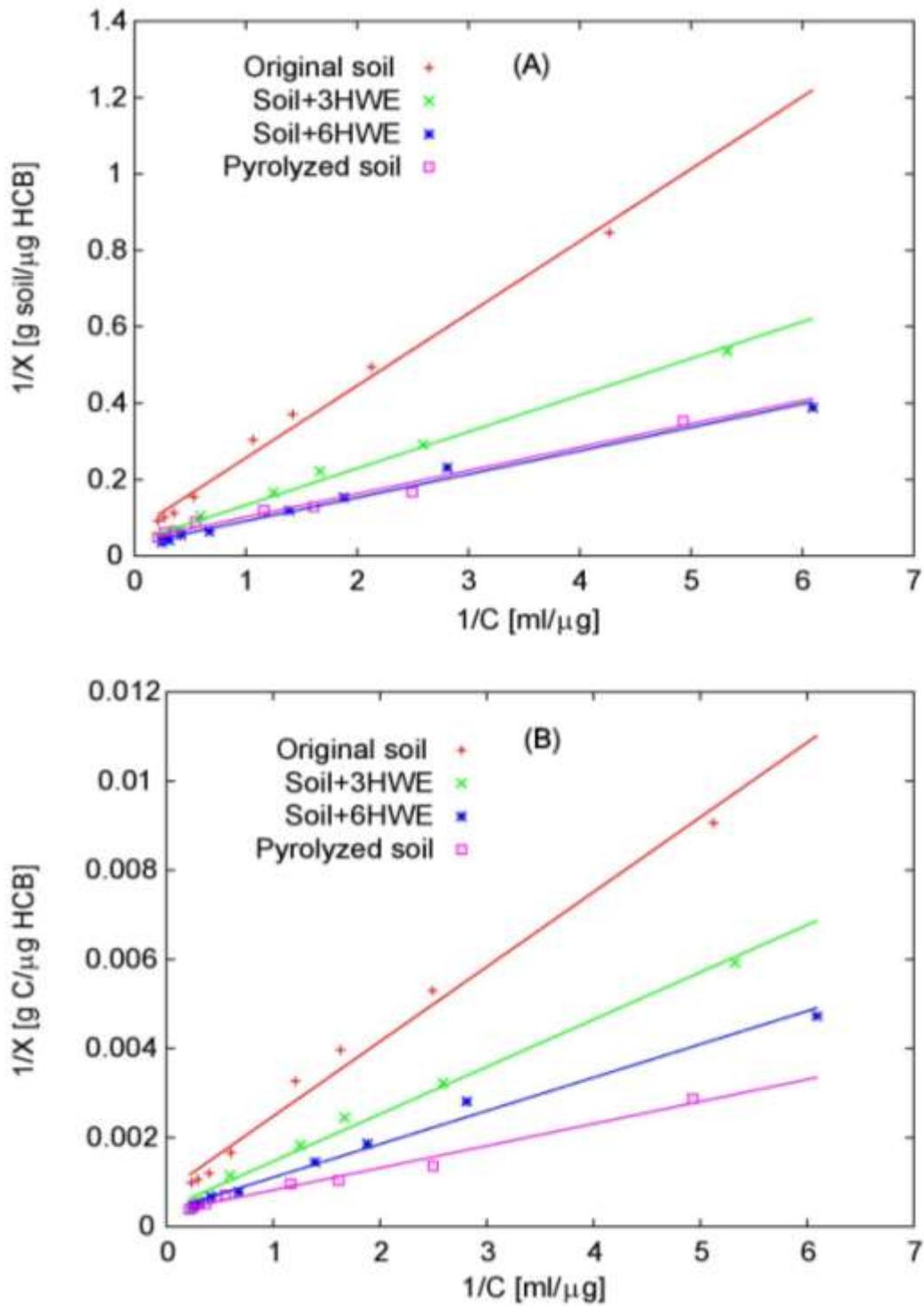

Figure 4. The Langmuir fitted adsorption isotherms of HCB on the soil samples normalized to the total soil mass, in μg HCB/g soil, (A) and the total carbon content, in μg HCB/g $C_{tot}$, (B).

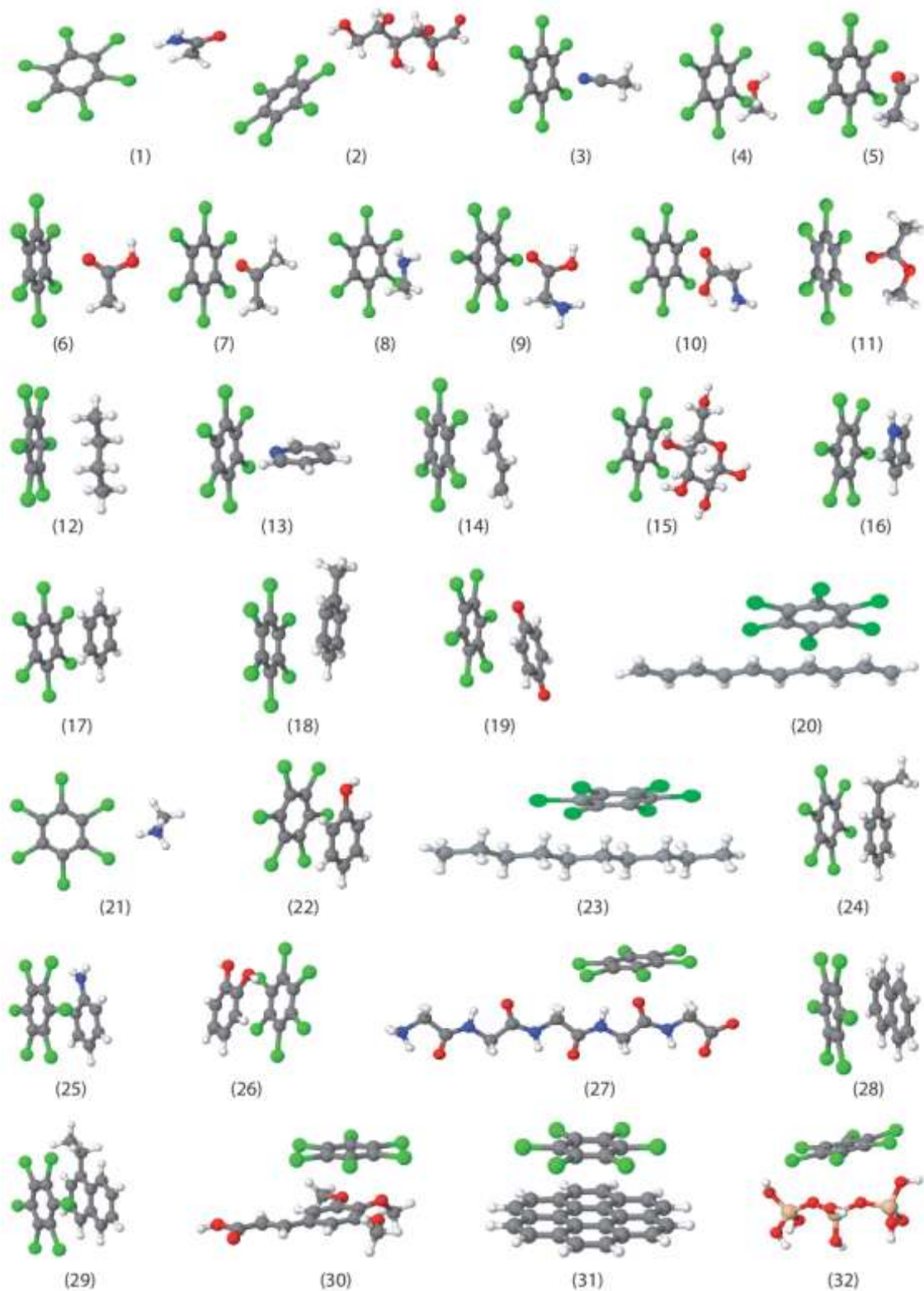

Figure 5. The test set developed for studying the interaction of HCB with different functional groups present in SOM. Shown are the gas phase geometries optimized using DFT/B3LYP and the 6-311++G(d,p) basis set together with the D3-Grimme correction for dispersion energies.

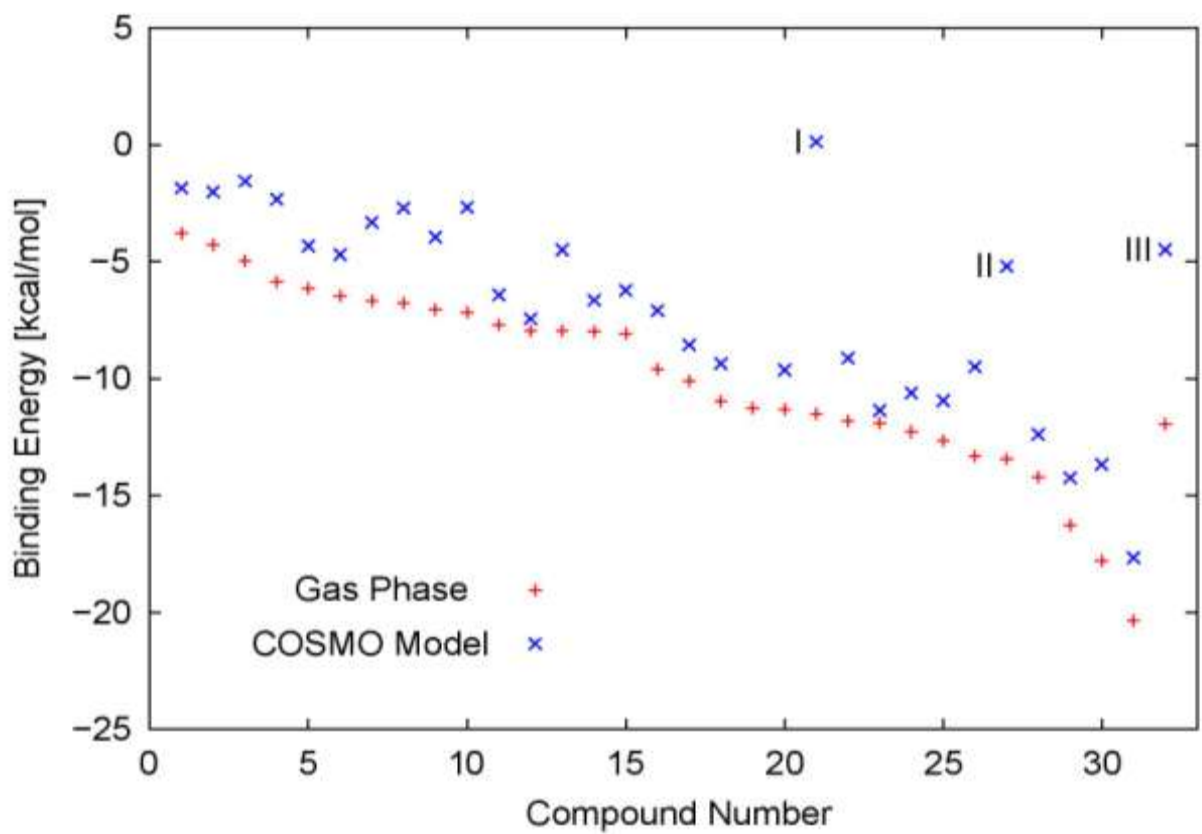

Figure 6. The binding energies for HCB with the test set given in Fig. 1 calculated at the B3LYP/D3/6-311++G(d,p) level of theory in gas phase and their corresponding binding free energies using the COSMO. (I), (II), and (III) mark complexes which are most strongly affected by solvation (21, 27, and 32).

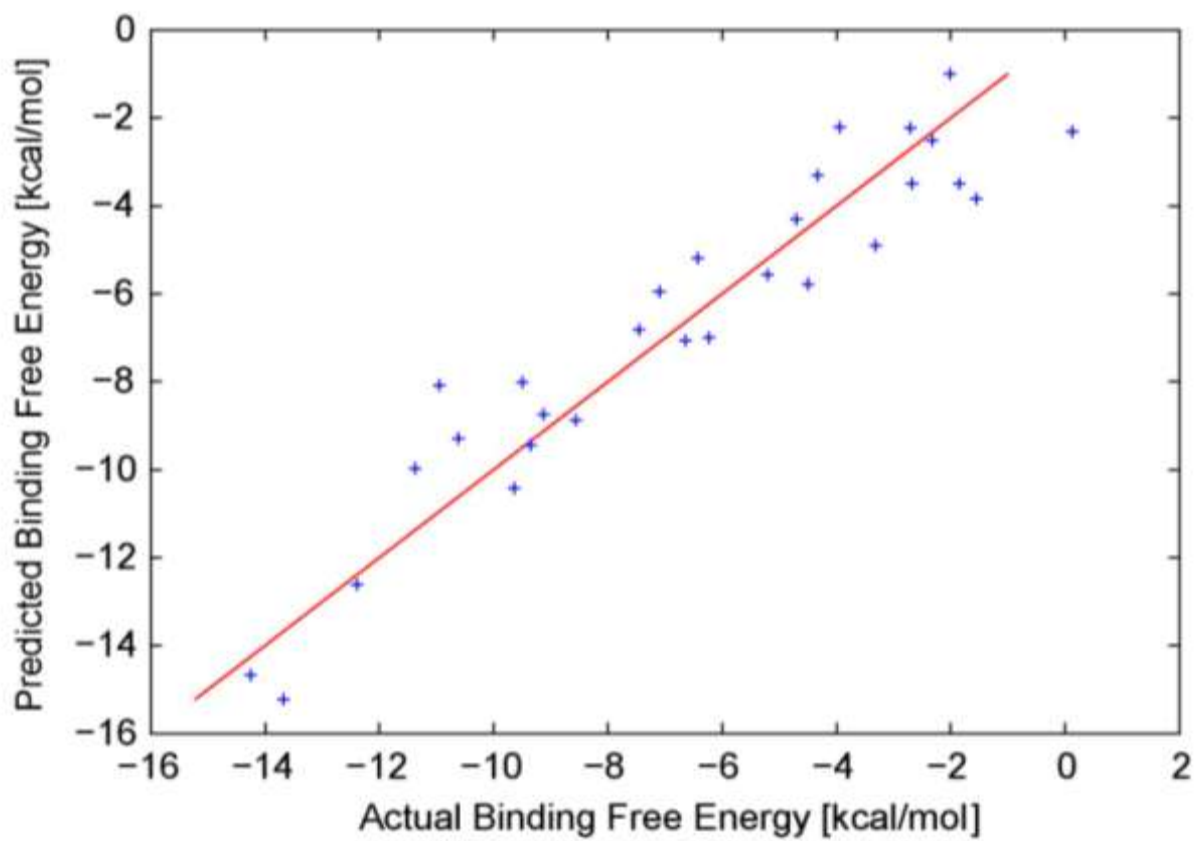

Figure 7. The predicted binding free energies of HCB with the SOM representative systems versus the calculated (actual) binding free energies at B3LYP/D3/6-311++G(d,p) and the red line is a linear correlation plotted as guide for eye.

Table 1. Fitted parameters with respect to Freundlich and Langmuir equations which are normalized to the total soil mass as well as the total carbon mass for the original soil, soil + 3 HWE, soil + 6 HWE, and the pyrolyzed soil samples.

| soil sample | Freundlich | | | | Langmuir | | | |
|---|---|---|---|---|---|---|---|---|
| | $K_F$ | | $n$ | $r^2$ | $K_L$ | $X_{max}$ | | $r^2$ |
| | total soil | $C_{tot}$ | | | | total soil | $C_{tot}$ | |
| original soil | 4.02 | 374.96 | 0.75 | 0.99 | 0.48 | 13.35 | 1245.78 | 0.99 |
| soil+3 HWE | 7.04 | 637.32 | 0.80 | 1.00 | 0.39 | 27.03 | 2445.47 | 0.99 |
| soil+6 HWE | 10.24 | 840.99 | 0.76 | 0.99 | 0.48 | 33.67 | 2764.42 | 0.99 |
| pyrolyzed soil | 8.66 | 1065.22 | 0.56 | 0.96 | 0.66 | 24.81 | 3051.57 | 0.98 |

Table 2. The absolute ion intensity in $10^4$ counts/mg of original soil (AII1), soil+3 HWE (AII2), soil+3 HWE (AII3); the absolute ion intensity difference between original soil and soil + 3 HWE (ΔAII1), and between soil + 3 HWE and soil + 6 HWE (ΔAII2); and the average binding energy (⟨$E_B$⟩) in kcal.mol-1 for the different SOM compound classes: phenols + lignin monomers = PHLM, alkyl aromatics = ALKY, carbohydrates = CHYDR, heterocyclic nitrogen containing compounds = NCOMP, peptides = PEPTI, lipids = LIPID, lignin dimers = LDIM (AII's as well as ΔAII's are adapted from (Ahmed et al., 2012))

| | PHLM | ALKY | CHYDR | NCOMP | PEPTI | LIPID | LDIM |
|---|---|---|---|---|---|---|---|
| *AII*1 | 74.2 | 76.8 | 34.2 | 30.3 | 12.7 | 25.9 | 14.9 |
| *AII*2 | 84.9 | 84.9 | 41.7 | 37.0 | 15.4 | 28.8 | 15.9 |
| *AII*3 | 86.1 | 86.1 | 44.4 | 42.2 | 18.3 | 32.2 | 18.3 |
| *ΔAII*1 | 10.7 | 8.1 | 7.5 | 6.7 | 2.7 | 2.9 | 1.0 |
| *ΔAII*2 | 1.2 | 1.2 | 2.7 | 5.2 | 2.9 | 3.4 | 2.4 |
| ⟨$E_B$⟩ | -10.8 | -11.0 | -6.2 | -5.8 | -5.2 | -8.8 | ----- |

Table 3. The fitted parameters, correlation of the binding energy with the relevant descriptors, and the most mutual correlated descriptors obtained using QSAR analysis (taking into account that +1.0 and -1.0 indicates to perfect positive and negative correlations, respectively, and 0.0 means that there is no any correlation).

| fitted parameter | value | correlated descriptor | correlation value | mutual descriptors | correlation value |
|---|---|---|---|---|---|
| SSR | 386.31 | $P_1$ | -0.89 | $P_2$ and $P_5$ | -0.98 |
| SSE | 45.90 | $P_2$ | 0.43 | $P_5$ and $P_6$ | 0.98 |
| SST | 432.21 | $P_3$ | 0.55 | $P_2$ and $P_6$ | -0.97 |
| $R^2$ | 89.38% | $P_4$ | -0.31 | $P_1$ and $P_6$ | 0.72 |
| adjusted $R^2$ | 86.49% | $P_5$ | -0.46 | $P_1$ and $P_5$ | 0.64 |
| $F_{statistic}$ | 30.86 | $P_6$ | -0.51 | $P_3$ and $P_4$ | -0.61 |
| critical F | 2.45 |  |  | $P_1$ and $P_2$ | -0.60 |